\documentclass[aps,prb,reprint,nofootinbib,twocolumn,superscriptaddress,showpacs,showkeys,longbibliography,amsmath,amssymb]{revtex4-2}
\usepackage{graphicx}
\usepackage{dcolumn}
\usepackage{bm}
\usepackage{braket}
\usepackage{subfigure}
\usepackage[colorlinks,bookmarks=false,citecolor=blue,linkcolor=red,urlcolor=blue]{hyperref}
\usepackage[english]{babel}

\begin{document}
\preprint{APS/123-QED}

\title{Adiabatic-impulse approximation in non-Hermitian Landau-Zener Model}
\thanks{A footnote to the article title}%
\author{Xianqi Tong}
\affiliation{Department of Physics, Beijing Normal University, Beijing 100000, People's Republic of China}
\author{Gao Xianlong}
\affiliation{Department of Physics, Zhejiang Normal University, Jinhua 321004, People's Republic of China}
\author{Su-peng Kou}\email[Corresponding author. Email: ] {spkou@bnu.edu.cn}
\affiliation{Department of Physics, Beijing Normal University, Beijing 100000, People's Republic of China}

\date{\today}

\begin{abstract}
We investigate the transition from PT-symmetry to PT-symmetry breaking and vice versa in the non-Hermitian Landau-Zener (LZ) models. The energy is generally complex, so the relaxation rate of the system is set by the absolute value of the gap. To illustrate the dynamics of phase transitions, the relative population is introduced to calculate the defect density in nonequilibrium phase transitions instead of the excitations in the Hermitian systems. The result shows that the adiabatic-impulse (AI) approximation, which is the key concept of the Kibble-Zurek (KZ) mechanism in the Hermitian systems, can be generalized to the PT-symmetric non-Hermitian LZ models to study the dynamics in the vicinity of a critical point. Therefore, the KZ mechanism in the simplest non-Hermitian two-level models is presented. Finally, an exact solution to the non-Hermitian LZ-like problem is also shown.
\end{abstract}

\pacs{Valid PACS appear here}

\maketitle

\section{Introduction}
The quantum two-level system exhibiting an avoided level crossing or level crossing plays an essential role in quantum adiabatic dynamics. If the control parameter is varied in time, the transition probability is usually captured by the Landau-Zener (LZ) theory \cite{LZ1, LZ2}. Usually, the quantum two-level system provides not only qualitative but also quantitative descriptions of system properties. It has become the standard theory for investigating many physical systems, e.g., the smallest quantum magnets, and Fe$_{8}$ clusters cooled below 0.36 K, are successfully described by the LZ model \cite{LZ2, LZ_model2}.

In many cases, the relevant parameters (i.e., the energy gap between the two levels on time) have the potential to be more general than the original LZ process. This motivates us to extend the level-crossing dynamics to level coalesce and various power-law dependencies in this paper. Appropriate changes in the external parameters driving the LZ transition can enable these LZ models to be experimentally realized in polarization optics \cite{polarization_optics1}, adiabatic quantum computing \cite{adiabatic_quantum_computing1,adiabatic_quantum_computing2}, and non-Hermitian photonic Lieb lattices \cite{photonic1,photonic2,photonic3}.

Fundamental axioms of quantum mechanics impose the Hermitian structure on the Hamiltonian. However, recent developments have shown the emergence of rich features for non-Hermitian Hamiltonians describing intrinsically non-unitary dynamics \cite{non-unitary1,non-unitary2,non-unitary3,non-unitary4,non-unitary5}, which have also been recently realized experimentally \cite{expriment1,NH_experiment1, NH_experiment2}. Although the eigenvalues of the non-Hermitian Hamiltonians can still be interpreted in terms of energy bands \cite{energy_bands1,energy_bands2}, the significance of their eigenvectors can no longer be handled by conventional methods because they are not orthogonal and thus already possess limited overlap without any additional perturbations \cite{NH_band1,NH_band2,NH_band3,NH_band4,NH_band5,NH_band6}. In this case, the exceptional points \cite{NH_EP2,NH_EP4,NH_EP5,NH_EP6,NH_EP7, NH_EP8} (EPs) are particularly important, where the complex spectra become gapless. These can be seen as non-Hermitian counterparts of the conventional quantum critical points \cite{NH_EP1, NH_EP3,convetional_CP1}. In EPs, two (or more) complex eigenvalues and eigenstates coalesce and then no longer form a complete basis \cite{NH_basis1,NH_basis2,NH_basis3}. Our main purpose is to study the linear quenching dynamics near the critical point, which is captured by the Kibble-Zurek (KZ) mechanism \cite{NH_EP4, NH_EP5, NH_KZ1, NH_KZ2, NH_KZ3, NH_KZ4, NH_KZ5, NH_KZ6}.

In this context, we present a successful combination of the KZ \cite{KZ1, KZ2,KZ3} theory of topological defect production and the quantum theory of the PT-symmetric non-Hermitian LZ model \cite{NH_LZ1, NH_LZ2, NH_LZ3, NH_LZ4}. Both theories play a prominent role in contemporary physics. The KZ theory predicts the production of topological defects (vortexes, strings) in the course of non-equilibrium phase transitions \cite{non_equilibrium1,non_equilibrium2,non_equilibrium3,non_equilibrium4,non_equilibrium5,non_equilibrium6,non_equilibrium7,non_equilibrium8}. This prediction applies to phase transitions in liquid $^{4}$\text{He} and $^{3}$\text{He}, liquid crystals, superconductors, ultra-cold atoms in optical lattices \cite{optical_lattice1,optical_lattice2}, and even to cosmological phase transitions in the early universe \cite{early_universe1,early_universe2}. To the best of our knowledge, the KZ mechanism in the simplest non-Hermitian two-level model has not been discussed before.

This work mainly focuses on the dynamical evolution of the PT-symmetric non-Hermitian LZ model, including adiabatic and impulse regimes during the slow quench of a system parameter. A real-to-complex spectral transition, which is usually called the PT transition, occurs in the non-Hermitian LZ model. In the PT-symmetric regime, the eigenvalues are real, ensuring the probability is conserved. When the energy gap is large enough away from the EP, the adiabatic theorem ensures that a system prepared in an eigenstate remains in an instantaneous eigenstate. This is in contrast to diabatic evolutions included by a very fast parameter change. 
This situation is more involved in the PT-broken regime where the eigenvalues are complex conjugates. Furthermore, the probability is no longer conserved because there is exponential growth and an exponential decay level, i.e., only the exponential growth state is left under the adiabatic evolution. Thus, the adiabatic conditions of the NH system were modified \cite{adiabatic_criterion1}. Near the EP, however, due to the reciprocal of the absolute value of the energy gap being greater than the change of parameters, the dynamics cannot be adiabatic, and the system gets excited. Then, since the modulus of the system is not conserved during the evolution, the relative occupation is proposed to calculate the excitation rather than the projection on the excited state. This scenario is captured by the adiabatic-impulse (AI) approximation.
Finally,  we also give non-trivial exact solutions for the non-Hermitian LZ model, successfully obtaining the theoretical free parameters in AI approximation.

In this paper, we want to illustrate the AI approximation from the simplest non-Hermitian two-level LZ model and discuss it in three different parts. Sect.~\ref{sec:symmetry} presents the adiabatic-impulse distinction in the PT-symmetric region, and the exact solutions of two quenching processes. In Sect.~\ref{sec:broken}, we also discuss the AI approximation solution of the PT-broken regime under different initial conditions. In Appendix~\ref{sec:Exact Expression for transition probability}, we explain the exact solution of the non-Hermitian Landau-Zener-like problem.  Details of analytic calculations are in Appendix~\ref{appendixa} (diabatic solutions in the PT-symmetric regimes).

\section{PT-symmetry}
\label{sec:symmetry}

\begin{figure}[]
	\centering
	\includegraphics[width=1.72in]{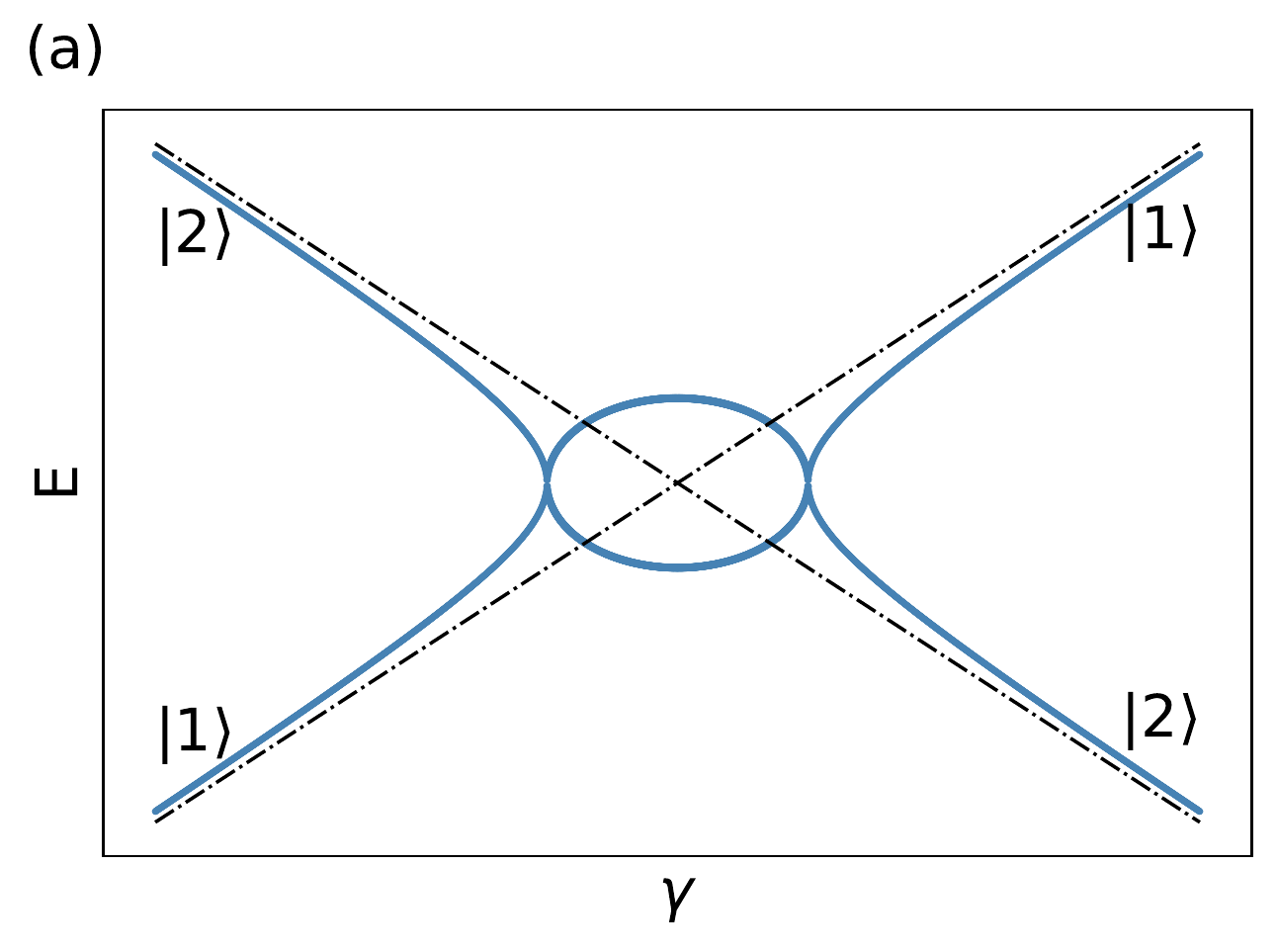}\includegraphics[width=1.71in]{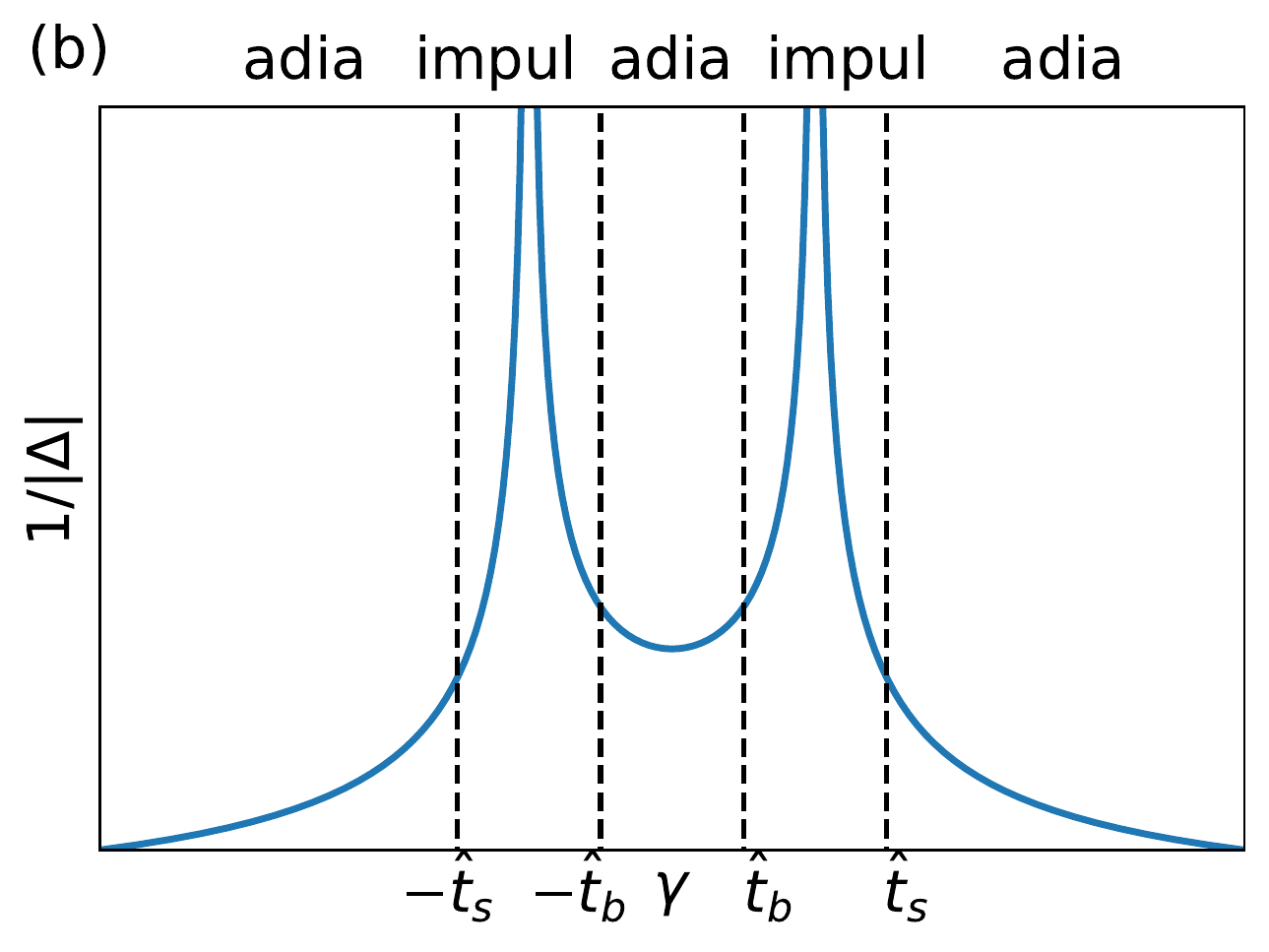}
\caption{(a) The energy level of the Hamiltonian (\ref{Hamiltonian_symmetry}); dotted line: $\nu=0$ case. Note the EP at $\gamma=\pm \nu\sqrt{\delta-1}$ and asymptotic form of eigenstates: $\ket{1},\ket{2}$. (b) The inverse of the energy gap in the non-Hermitian LZ model. The four dashed lines correspond to the instants in the PT-symmetry and PT-symmetry breaking regimes, which separate the adiabatic and impulse regimes.}
\label{fig:phase diagram}
\end{figure}
The PT-symmetric non-Hermitian LZ model we consider is
\begin{equation}
    H(t)=\frac{1}{2}\left( \begin{array}{cc}
            (-1)^{n} \gamma & \nu \\
            \nu (1-\delta) & (-1)^{n+1} \gamma
            \end{array} \right),
     \label{Hamiltonian_symmetry}
\end{equation}
where the $\gamma=\Delta t$ is time-dependent, $\Delta$ is a time-independent constant, and $n=0$ in this part. The system experiences adiabatic time evolution when $\Delta \rightarrow 0$, and $\Delta \rightarrow \infty$ means diabatic evolution. In this model, $\nu$, $\delta>0$ are constant parameters. We set $\nu=1$ as an energy unit that does not influence the results.

For a non-Hermitian Hamiltonian $H$, let $\ket{ i^{R} }$ denote the $i$th left eigenstates with the (generally complex) eigenenergy $E_{i} $, i.e., $\left\langle i^{L}\right| H =\left\langle i^{L} \right| E_i$. Note that the $m$th right eigenvector $\ket{j^{R} }$ satisfies $H \left|j^{R} \right\rangle=E_j \left| j^{R} \right\rangle$ and the biorthonormal relation $\braket{i^{L} | j^{R}} = \delta_{ij}$.

At any instantaneous time, the right eigenstates of this Hamiltonian can be written on the time-independent basis $\ket{1}$ and $\ket{2}$. The ground state $\ket{\downarrow(t)^{R}}$ and the excited state $\ket{\uparrow(t)^{R}}$ are given by the following equation,
\begin{equation}
    \left[\begin{array}{l}
|\uparrow(t)^{R} \rangle\\
|\downarrow(t)^{R} \rangle
\end{array}\right]=\left(\begin{array}{cc}
-\frac{i}{\sqrt{\delta-1}}\cosh\frac{\theta}{2} & i\sinh\frac{\theta}{2}\\
-\frac{1}{\sqrt{\delta-1}}\sinh\frac{\theta}{2} & \cosh\frac{\theta}{2}
\end{array}\right)\left[\begin{array}{l}
|1 \rangle\\
|2 \rangle
\end{array}\right],
\end{equation}
where $\cosh \theta=\epsilon/\sqrt{\epsilon^{2}-1}$, $\sinh \theta=1/\sqrt{\epsilon^{2}-1}$, $\epsilon=\gamma/\nu\sqrt{\delta-1}$. We consider only $\theta \in [0, \pi]$. If $\theta$ is not a real number, the system's PT symmetry is broken, and this part is not considered. The energy spectrum is depicted in Fig. \ref{fig:phase diagram} with the energy gap $\Delta_{g}=\sqrt{\gamma^{2}+\nu^{2}(1-\delta)}$. It can be seen that the energy gap $\Delta_{g}=0$ at the exceptional points $\gamma_{EP}=|\nu \sqrt{\delta-1}|$, which is accompanied by the coalesce of eigenvalues and eigenstates.

The density of topological defects can be introduced in the non-Hermitian LZ model in the following way. Suppose the state $\ket{1}$ is the eigenstate of arbitrary non-Hermitian operate $\hat{O}$: $\hat{O} \ket{1^{R}}=n \ket{1^{R}}$ $(n=\pm1, \pm2, ...)$, while the state $\ket{2^{R}}$ always corresponds to the 0 eigenvalue, i.e. $\hat{O} \ket{2^{R}} = 0 \ket{2^{R}}$. So for any normalized state can be written as $\ket{\Psi}= a\ket{1^{R}}+ b \ket{2^{R}}$ ($|a|^{2}+|b|^{2}=1, \braket{i ^{L} | j ^{R}}= \delta_{ij} $ in the PT-symmetry regime). The unity density defect is determined by the expectation value of operator $\hat{O}$, $\overline{\braket{ \hat{O} }}=\braket{\Psi|\hat{O}|\Psi}/n = \left| \braket{\Psi|1} \right|^{2}$. But for non-unitary evolution in the PT-broken area, the probability of time evolution state on the instantaneous states is not conserved, i.e., $|a|^{2}+|b|^{2}\neq 1$. Then,  $\overline{\braket{O}}$ is replaced with the relative occupation
\begin{equation}
    \mathcal{D}_{r}=\frac{\left| \braket{1^{L}|\Psi} \right|^{2}}{ \left| \braket{1^{L}|\Psi} \right|^{2}+ \left| \braket{2^{L}|\Psi} \right|^{2}},
    \label{eq:relative occupation}
\end{equation}
where $\ket{n^{L}}$ is the $n$th left eigenstate. When we discuss only the PT-symmetric regime, where $\left| \braket{1^{L}|\Psi} \right|^{2}+ \left| \braket{2^{L}|\Psi} \right|^{2}=1$, $\mathcal{D}_{r}$ returns to the Hermitian case, i.e., $\mathcal{D}_{r}=\overline{\braket{\hat{O} }} $.

Suppose the system evolves adiabatically from the ground state of (\ref{Hamiltonian_symmetry}) at $| t | \rightarrow \infty$ to the ground state across the EP. Therefore, the state of the system will go from a density-free phase to a density-defected one, that is, undergo a phase transition from $\ket{1^{R}}$ to $\ket{2^{R}}$. If the time evolution fails to be adiabatic, which is usually the case, the state at the end is a superposition of states $\ket{1^{R}}$ and $\ket{2^{R}}$, so the expected value of the operator $\hat{O}$ is nonzero. Then we will show that the KZ theory can well predict the topological density (\ref{eq:relative occupation}) in the non-Hermitian LZ system.

The analogy of relaxation time scale, relative temperature, and quench time scale are determined as follows. First, the KZ theory neatly simplifies the evolution of system dynamics. The simplification is the essence of the KZ mechanism that suggests splitting the quench into the regime near the EP and the quasiadiabatic regime far from the EP, that is to say, the state becomes changeless (impulse), or can adjust to changes in the parameter. This is the key concept from Zurek \cite{zurek1,zurek2,zurek3}, and the switch from adiabatic to impulse is determined by the relevant time scale. And the relevant time scale equals the reciprocal of the energy gap, which is small when the parameter is apart from the EP and relatively large in the impulse regime. According to the adiabatic theorem, as long as the reciprocal of the gap is small enough, the system will evolve from the ground state and remain in the ground state. This naturally shows that the reciprocal of the gap must be small in the adiabatic evolution regime, which can be regarded as the equivalent relaxation time scale introduced above: $\tau=1/\sqrt{\gamma^{2} + \nu^{2} (1-\delta) }$. The dimensionless distance $\epsilon=\Delta t/ \left(\nu \sqrt{\delta-1}\right)$ of the system from the exceptional point is the relative temperature. The quench time is $\tau_{Q}=\nu \sqrt{\delta-1} /\Delta$. Then, $\nu \sqrt{\delta-1}$ is identical with $1/ \tau_{0}$. Finally, the relaxation time can be rewritten as
\begin{equation}
    \tau=\frac{\tau_{0}}{\sqrt{\epsilon^{2}-1}}, \quad \epsilon=\frac{t}{\tau_{Q}}.
    \label{relaxation time}
\end{equation}
For $\left| \epsilon \right| \gg 1$, the relaxation time $\tau \approx \tau_{0}/ |\varepsilon|$ will be back to topological defect density in liquid $^{4}$He \cite{zurek1,zurek2,zurek3}, which will be discussed in details below.

In the following, we will consider the dynamics of the LZ model described by the time-dependent Schr\"{o}dinger equation $i \frac{d}{d t}|\Psi\rangle=\hat{H}(t)|\Psi\rangle$, see Eq. (\ref{Hamiltonian_symmetry}). When the whole evolution begins at time $t_{i}=-\infty$, the initial state is chosen to be the ground state $ \ket{ \phi_{{G}} }$, and lasts till $t_{f} \rightarrow -\gamma_{EP}$. Since the eigenvalues coalesce at EP, no matter how slowly the parameters are driven, it is impossible for the system's quantum state to evolve adiabatically near the critical point. This paper aims to quantify this unavoidable excitation level in non-Hermitian systems. At the beginning of the evolution, the energies are real, and the energy gap is large enough so that the states of the system evolve adiabatically. On the contrary, when the time-dependent parameter of the Hamiltonian is gradually approaching an exceptional point, the time-evolved state can not follow the change of the parameter of the Hamiltonian. The evolved state becomes an impulse near the EP. Furthermore, when the system is in the PT-broken regime, only one state dominates the population, i.e., the eigenstate with the largest imaginary eigenvalue. So, the whole evolution stage can be divided into two different regimes in PT-symmetric or (PT-broken) regime:
\begin{equation}
\begin{aligned}
   &t \in(-\infty,-\hat{t}):|\Psi(t)\rangle \approx \text { (phase factor) } \ket{ \phi_{{G}}(t)}, \\
   &t \in [ -\hat{t}, -\gamma_{EP}]:|\Psi(t)\rangle \approx \text {(phase factor)} \ket{ \phi_{{G}}(-\hat{t})}, \\
\end{aligned}
\label{Schematic diagram}
\end{equation}
which is same for the evolution $t \in [ \gamma_{EP},+\infty)$, because the energy spectrum is real and symmetric on both sides when $\left| \gamma \right|>\gamma_{EP}$.

According to Eq. (\ref{Schematic diagram}), the state will be impulsed in the regime near the EP that only has a different phase factor. And if the state can adiabatically evolve during the time, we can approximately consider the state as the instantaneous eigenstate of the Hamiltonian with a different phase factor. Of course, the process will return to the adiabatic evolution when the real energy gap is big enough again. The assumption behind Eq. (\ref{Schematic diagram}) above is based on how well the KZM works in the non-Hermitian LZ model.

However, the instants $\pm \hat{t}$ are still unknown. It is firstly calculated from the equation proposed by Zurek in the paper on classical phase transition \cite{zurek1,zurek2,zurek3}
\begin{equation}
    \tau(\hat{t})=\alpha \hat{t},
    \label{AI condition}
\end{equation}
and $\alpha=O(1)$ is a constant $independent$ of $\tau_{{Q}}$ \cite{KZ1,KZ2}. In the PT-symmetry regime, the solution of Eq. (\ref{AI condition}) reads 
\begin{equation}
    \hat{\varepsilon}_{s}=\varepsilon_{s}(\hat{t})=\frac{1}{\sqrt{2}} \sqrt{\sqrt{1+\frac{4}{x_{\alpha}^{2}}}+1}, \quad x_{\alpha}=\alpha \frac{\tau_{Q}}{\tau_{0}}.
\end{equation}
For fast transition, i.e., $x_{\alpha} \rightarrow 0$, we get $\hat{t}=\sqrt{\tau_{0} \tau_{Q} / \alpha}$.

Then, the first case we consider is completely in the PT-symmetric region. The initial state $\ket{\Psi(t_{i})}$ is set to ground state $\ket{\downarrow(t_{i})}$, and the start evolution point away from the EP, i.e., from $-\infty$ to $-\gamma_{EP}$. Thus, we can calculate the occupation of the final state on the left eigenstate, rather than the right eigenstate in the Hermitian system:
\begin{figure}
    \centering
    \includegraphics[scale=0.62]{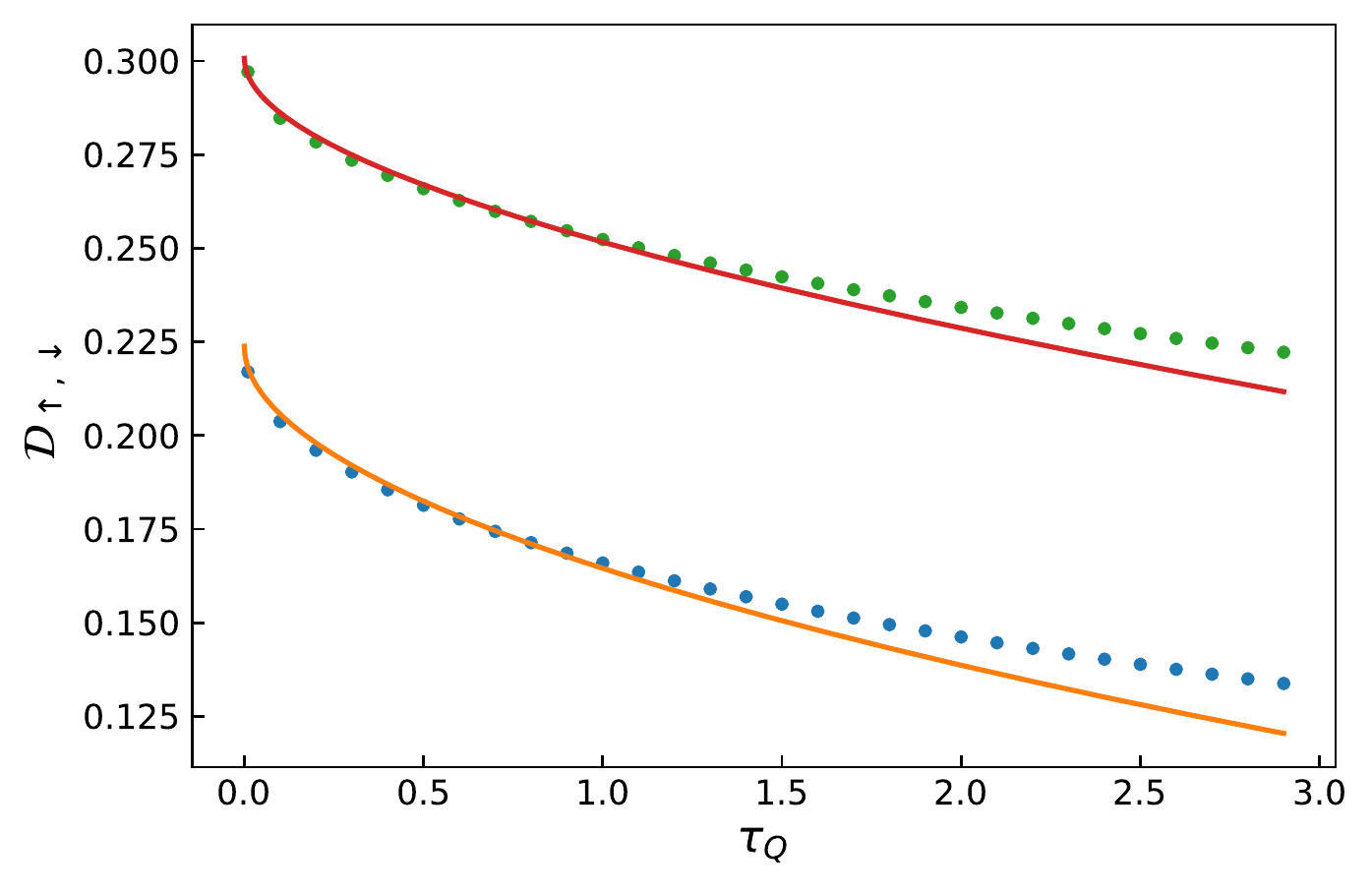}
    \caption{Transition probability as a function of $\tau_{Q}$ in the PT-symmetric regime for two different: $\mathcal{D}_{\uparrow}$ (upper curves), $\mathcal{D}_{\downarrow}$ (lower curves). Dots: numerical data by solving the Schr\"{o}dinger equation, solid lines: AI approximation (\ref{eq:PT_density_1}), (\ref{eq:PT_density_3}) in which the $\alpha=0.06$ is determined from the exact solution of Appendix \ref{sec:Exact Expression for transition probability}. Upper (lower) curves all corresponding to $\theta_{0}=0.50\pi$ $(0.01\pi)$.}
    \label{fig:PT_near_ep}
\end{figure}
\begin{equation}
\begin{aligned}
	\mathcal{D}_{\downarrow} &=\left|\left\langle \downarrow(t_{f})^{L} \mid \widetilde{\Psi}\left( t_{f} \right)\right\rangle\right|^{2}\\
	&\approx |\langle\downarrow(t_{f})^{L} \mid \uparrow(-\hat{t})^{R}\rangle|^{2}\\
	&=\frac{1}{ -\cosh\theta_{0}\sqrt{2P(x_{\alpha})}-2\sinh\theta_{0}\sqrt{P(x_{\alpha})/2-1}}+\frac{1}{2},
	\label{eq:PT_density_1}
\end{aligned}
\end{equation}
where $\theta_0=\arctan \left(\tau_{Q} /t_{i}\right) \in[-\pi, \pi].$ measures the distance between the start or end point of the time evolution and the critical point, $P\left(x_{\alpha}\right)=2+x_{\alpha}^{2}+x_{\alpha}\sqrt{4+x_{\alpha}^{2}}$, and the $\ket{\widetilde{\Psi}\left(t_{f}\right)}$ is the normalized time-dependent state. During the whole evolution of the non-Hermitian system, we normalize the wave function at every step $dt$. A derivation of Eq. (\ref{eq:PT_density_1}) above is $| \braket{\downarrow(t_{f})^{L} \mid \widetilde{\Psi}\left( t_{f} \right)}|^{2} \approx |\langle\downarrow(t_{f})^{L} \mid \widetilde{\Psi}(-\hat{t})\rangle|^{2} = |\langle\downarrow(t_{f})^{L} \mid \uparrow(-\hat{t})^{R}\rangle|^{2}$. 

In addition, we do not test the case of cross EP, because the initial state information will be erased after EP, where two eigenvectors coalesce. Even at the adiabatic limit ($\tau_{Q}\rightarrow\infty$), the $\mathcal{D}_{\downarrow}$ of the final state at the EP is almost the same.

The excitation probability $\mathcal{D}_{\downarrow}$ will be expanded into a series for fast transitions
\begin{align}
	\mathcal{D}_{\downarrow}&=\sinh ^2\left(\frac{\text{$\theta_{0}$}}{2}\right) \text{sech}(\text{$\theta_{0}$})  +\frac{1}{2} \tanh (\text{$\theta_{0}$}) \text{sech}(\text{$\theta_{0}$})\sqrt{x_{\alpha}} \notag\\
	&-\frac{1}{8} (\cosh (2 \text{$\theta_{0}$})-3) \text{sech}^3(\text{$\theta_{0}$})x_{\alpha}+O(x_{\alpha}^{3/2}).
\end{align}
Here the first two terms are equal to $0$ at $t_{i} \rightarrow -\infty$, which are trivial. The coefficient in front of $x_{\alpha}$ is $\underset{\theta_{0}\rightarrow0}{lim}-\frac{1}{8} \left(  \text{sech}^3(\text{$\theta_{0}$}) (\cosh (2 \text{$\theta_{0}$})-3)\right)=\frac{1}{4}$. Then, we arrive
\begin{equation}
	\mathcal{D}_{\downarrow}=\frac{1}{4}\alpha\tau_{Q}+O(\tau_{Q}^{3/2}).
\end{equation}
The second case is the initial state is the ground state at $t_{i}=\tau_{Q}/ \tan(\pi/2)$, ending at $t_{f} \rightarrow \infty$. As discussed in Eq. (\ref{eq:PT_density_1}), by using the AI approximation, we can easily derive the following predictions, that is, the probability of finding the system in an excited state at $t \rightarrow \infty$:
\begin{align}
	\mathcal{D}_{\uparrow} &= | \braket{\uparrow(t_{f})^{L} \mid \widetilde{\Psi}\left( t_{f} \right)}|^{2} \notag\\
	&\approx |\langle\uparrow(\hat{t})^{L} \mid \uparrow(\hat{t})^{R}\rangle|^{2} \notag\\
	 &=\frac{1}{\cosh\theta_{0}\sqrt{2P(x)}-2\sinh\theta_{0}\sqrt{P(x)/2-1}}+\frac{1}{2}.
	 \label{eq:PT_density_3}
\end{align}
We assumed that $|\braket{\uparrow(t_{f})^{L} |\tilde{\Psi}(t_{f})} |^{2} \approx \mid\langle\uparrow(\hat{t})^{L} | \widetilde{\Psi}(\hat{t})\rangle|^{2} \approx$ $\mid \langle\uparrow(\hat{t})^{L} | \widetilde{\Psi}(t_{i}) \rangle|^{2} = \mid\langle\uparrow(\hat{t})^{L} | \downarrow(t_{i})^{R}\rangle|^{2}$.

Expand $\mathcal{D}_{\uparrow}$ into a series for a fast transformation
\begin{align}
	\mathcal{D}_{\uparrow}&=\sinh ^2\left(\frac{\text{$\theta_{0}$}}{2}\right) \text{sech}(\text{$\theta_{0}$})-\frac{1}{2} \tanh (\text{$\theta _{0}$}) \text{sech}(\text{$\theta_{0}$}) \sqrt{x_{\alpha}} \notag \\
	&+O(x_{\alpha}),
\end{align}
where the $\theta_{0}\rightarrow \pi/2$. Determination of the constant $\alpha$ is presented in Appendix~\ref{sec:Exact Expression for transition probability}, by substituting $\eta=1/2$ into Eq. (\ref{eq:symmetry_density3}), we can get $\alpha=\frac{\left(1+e^{\pi }\right) \pi}{8 \left(e^{\pi /2}-1\right)^4} \left(1-2 S\left(\frac{\coth \left(\frac{\pi }{2}\right)}{\sqrt{\pi }}\right)\right)^2$, where $S$ is the Fresnel integral $S(z)=\int _0^z d t \left( \sin\left(\pi  t^2/2 \right) \right)$.

A comparison of  Eq. (\ref{eq:PT_density_1}) and Eq. (\ref{eq:PT_density_3}) to the numerical results for $\tau_{Q}/\tau_{0}<1.75$ and $|\theta_{0}-\pi/2| \leq \pi/10$ proclaims satisfactory agreement; (see Fig. \ref{fig:Pb_6}). For larger $\tau_{Q}$ or $|\theta_{0}-\pi/2|$ the agreement gradually decreases, due to the fact that for the parameter, $t_{i}=\tau_{Q}/\tanh(\theta_{0})$, maybe outside the impulse regime $[-\hat{t},\hat{t}]$, where the initial assumption is violated. One avoids these problem when either $t_{i}\ll -\hat{t}$ or $|t_{i}|\ll \hat{t}$, i.e., when the entire evolution of the system is clearly divided into adiabatic and frozen parts.

\section{PT-broken}
\label{sec:broken}

\begin{figure}[]
\centering
\includegraphics[width=0.48\textwidth]{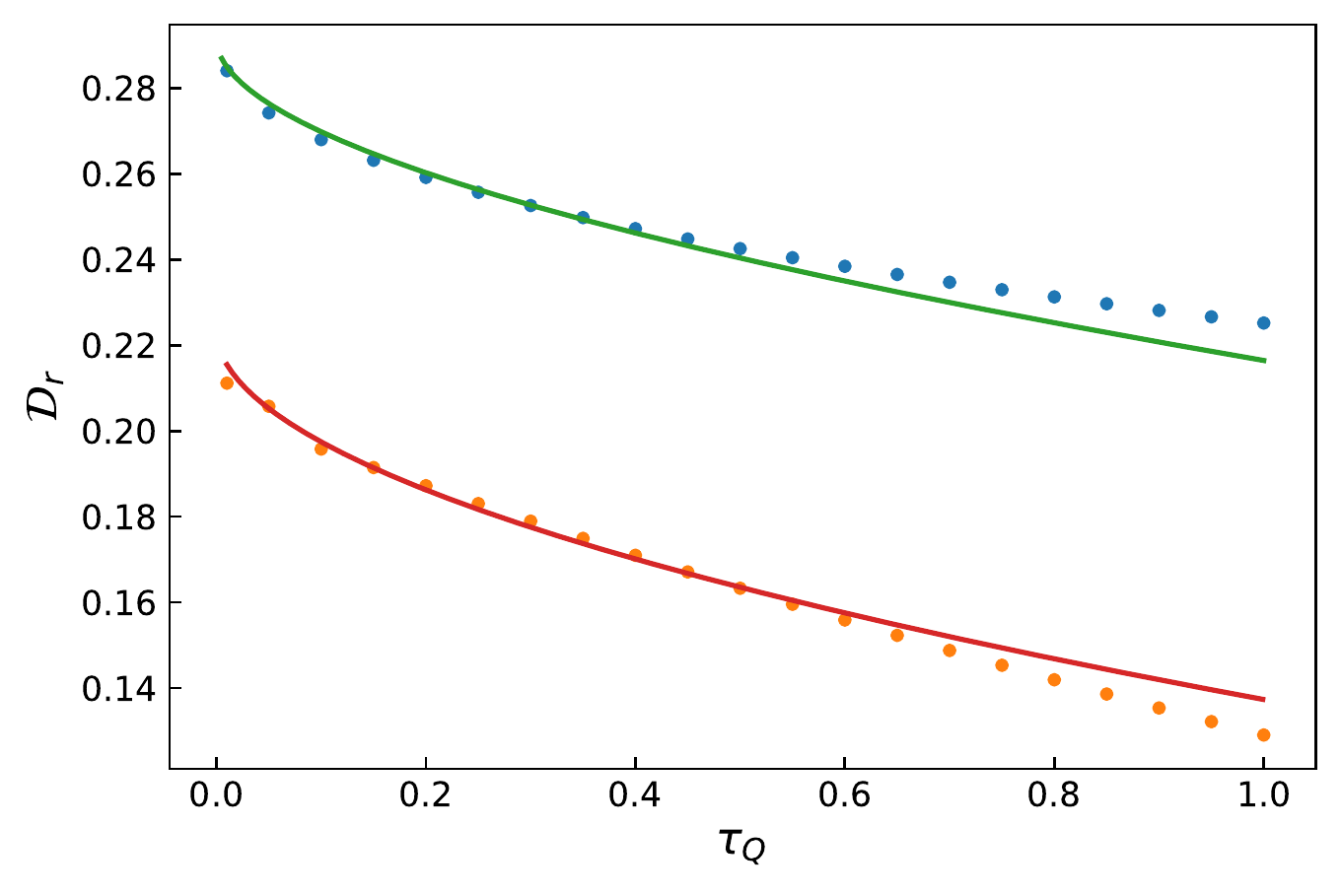}
\caption{Transition probability in the PT-broken regime. Dot: numerics by solving the Schr\"{o}dinger equation, solid lines: AI approximation (\ref{eq:Pb_near_ep}) in which $\alpha$ is determined from the exact solution in Appendix~\ref{sec:Exact Expression for transition probability}. Upper (lower) curves correspond to $\theta_{0}=1.25\pi$ $(0.2\pi)$, while $\alpha \approx 0.12$ is the same.}
\label{fig:Pb_6}
\end{figure}
When the evolution is a full non-Hermitian drive, can we see similar relation between excitation and quenching time in the PT-broken region? In the Hamiltonian (\ref{Hamiltonian_symmetry}), we assume $n=1/2$, $\sqrt{-1}= i$ is a imaginary number. And the $\delta$ always be set to 0 in the PT-broken regime. This full non-Hermitian drive is equivalent to quenching the imaginary tachyon mass \cite{expriment1}. Then the eigenvalues of the Hamiltonian are
\begin{equation}
	E_{\pm}=\pm \frac{i}{2}\sqrt{\gamma^{2}-\nu^{2}}.
\end{equation}
When $|\gamma|>\nu$, the system is in the PT-broken regime, and all eigenvalues are imaginary. Otherwise, the eigenvalues are all real numbers when $|\gamma|<\nu$, i.e., the system is in the PT-symmetric regime.

Since the eigenvalues are complex conjugates, one of the projections of the time-dependent wave function on the left eigenstate is exponentially increasing, and the other is exponentially decayed. Likewise, we assume that time starts from the ground state of $-\infty$ and continues to $-\gamma_{EP}$, or from $\gamma_{i}$ to $\infty$, where $\gamma_{i}$ is in the impulse regime on the other side. However, the ground state here refers to the eigenstate corresponding to the positive eigenvalue because, if $\gamma$ changes adiabatically, the system has enough time to evolve to the exponentially growing state. This state corresponds to the least-dissipative instantaneous eigenstate with the largest imaginary eigenvalue and dominates the adiabatic process. Thus, the initial state is always chosen to be the least-dissipative eigenstate $\ket{ \phi_{{l}}(t_{i})^{R}}$. When the time-dependent state evolves to the vicinity of the exception point, the state will be excited. The density of topological defects calculated by the relative population is a function of $\tau_{Q}$. Interestingly, we found that the KZ mechanism is still valid in the broken regime, which can describe the slow quench dynamics well. The KZ mechanism only, however, describes the density in the impulse regime when the instantaneous transition rate $\dot{\epsilon}/\epsilon$ is much larger than the energy gap $\Delta_{g}$. Therefore, these two important options can be considered.
\begin{equation}
\begin{aligned}
   &t \in(-\infty,-\hat{t}):|\Psi(t)\rangle \approx \text { (phase factor) } \ket{ \phi_{{l}}(t)^{R} }, \\
   &t \in [ -\hat{t}, -\gamma_{EP}]:|\Psi(t)\rangle \approx \text {(phase factor)} \ket{ \phi_{{l}}(-\hat{t})^{R}}. \\
\end{aligned}
\label{Schematic_diagram_broken}
\end{equation}
The same holds in the $t\in [\hat{t}, \infty )$ interval. Adiabaticity in the Hermitian system means that the probabilities are constant. But in the non-Hermitian systems, only the least dissipative state can evolve adiabatically, because the probabilities on the other states are suppressed. In the freezing regimes, however, the definition is the same as before.

However, the eigenvalues are pure imaginary numbers, so the ``relaxation time" is not the energy gap of real eigenvalues but rather the absolute value of the imaginary eigenvalue difference
\begin{align}
\tau=\frac{\tau_{0}}{ | \sqrt{1-\epsilon^{2}} |}, \quad \epsilon=\frac{t}{\tau_{Q}},
\label{relaxation time_broken}
\end{align}
where $\tau_{0}=\nu=1$. Naturally, after bringing Eq. (\ref{relaxation time_broken}) into Eq. (\ref{AI condition}), one obtains the dimensionless distance $\hat{\varepsilon}_{b}=\frac{1}{\sqrt{2}} \sqrt{1+\sqrt{1+\frac{4}{x_{\alpha}^{2}}}} =\hat{\varepsilon}_{s}$.

Now, we consider the case when the initial state is chosen to the least eigenstate prepared at impulse area closing to the EPs, and assume the limits $t_{f} \rightarrow \hat{t}$. Taking the initial state $\left|\Psi\left(t_{i}\right)\right\rangle=\left|\uparrow\left(t_{i}\right)\right\rangle:=\frac{1}{\sqrt{1-e^{-2 \alpha_{0}}}} i e^{\alpha_{0}} |1\rangle+ \frac{1}{\sqrt{1-e^{-2 \alpha_{0}}}} |2\rangle$, where $\alpha_{0}=\text{arccosh} \left|\gamma_{i} \right| \in [0, \infty )$. However, due to the non-unitary time-dependent evolution, $\mathcal{D}_{\uparrow}$ may be greater than 1. Instead of the $\mathcal{D}_{\uparrow}$, the relative population of the instantaneous eigenstates is proposed to calculate the density \cite{NH_LZ2}:
\begin{align}
	\mathcal{D}_{r} &=\frac{\left|\braket{\uparrow(t_{f})^{L} |\tilde{\Psi}(t_{f})}\right|^{2}}{\left|\braket{\uparrow(t_{f})^{L}|\tilde{\Psi}(t_{f})}\right|^{2}+\left|\braket{\downarrow(t_{f})^{L}|\tilde{\Psi}(t_{f})}\right|^{2}} \notag\\
	&\approx\frac{\mid\langle\uparrow(\hat{t})^{L} | \downarrow(t_{i})^{R}\rangle|^{2}}{\mid\langle\uparrow(\hat{t})^{L}| \downarrow(t_{i})^{R}\rangle |^{2}+\mid\langle\downarrow(\hat{t})^{L}| \downarrow(t_{i})^{R}\rangle |^{2}} \notag\\
	&=\frac{1}{2}\frac{\sinh(\alpha_{0})}{\sqrt{P(x_{\alpha})/2-1}-\sqrt{P(x_{\alpha})/2}\cosh(\alpha_{0})}+\frac{1}{2},
	\label{eq:Pb_near_ep}
\end{align}
where $\alpha_{0}=\text{arccosh} \left|t_{i}/\tau_{Q} \right|$ measures the distance of the starting point from the exceptional point, e.g., $\alpha_{0}=\pi/2$ when evolution starts from an crossing center. And assumed that $\left|\braket{\uparrow(t_{f})^{L} |\tilde{\Psi}(t_{f})}\right|^{2} \approx \mid\langle\uparrow(\hat{t})^{L} | \widetilde{\Psi}(\hat{t})\rangle|^{2} \approx \mid\langle\uparrow(\hat{t})^{L} | \widetilde{\Psi}(t_{i})\rangle|^{2} = \mid\langle\uparrow(\hat{t})^{L} | \downarrow(t_{i})^{R}\rangle|^{2}$. However, in the PT-broken phase, it is worth noting that truncation is needed for time $t_{f}$ to ensure that the probability is not greater than 1, that is, when $\dot{\epsilon}/\epsilon \ll \Delta_{g}$.

AI predictions can be easily compared to accurate results Eq. (\ref{eq:broken_density3}) after series expansion
\begin{align}
	\mathcal{D}_{r}=\frac{1}{e^{2 \alpha _0}+1}+\frac{\tanh \left(\alpha _0\right) \text{sech}\left(\alpha _0\right)}{2} \sqrt{x_{\alpha}}+O\left(x _{\alpha}^{1/2}\right),
	\label{eq:AI_broken_density3_series}
\end{align}
where the $\alpha$ is easily determined and is presented in Appendix~\ref{sec:Exact Expression for transition probability}. By putting $\eta=1/2$ into Eq. (\ref{eq:broken_density3}) one can get $\alpha \approx 0.12$.

Then we consider the case that evolves the system to the vicinity of the critical point, $t_{f}\rightarrow t_{n}$ while starting from the ground state at $t_{i} \rightarrow -\infty$. 
\begin{align}
	&\left|\braket{\downarrow(t_{f})^{L}| \widetilde{\Psi}( -\hat{t}) }\right|^{2}\notag\\
	&\approx|\langle\downarrow( t_{f})^{L} \mid \widetilde{\Psi}( -\hat{t})\rangle|^{2} \approx \mid\langle\downarrow(t_{f})^{L} | \downarrow(-\hat{t})^{R} \rangle|^{2}.
	\label{eq:AI_broken_density1_series}
\end{align}
One easily gets the final density of topological defects the same as Eq. (\ref{eq:Pb_near_ep}). Substituting  Eq. (\ref{eq:AI_broken_density1_series}) into Eq. (\ref{eq:relative occupation}), the excitation is obtained after the expansion equal to
\begin{align}
	\mathcal{D}_{r}=\frac{1}{e^{2 \alpha _0}+1}-\frac{\tanh \left(\alpha _0\right) \text{sech}\left(\alpha _0\right)}{2} \sqrt{x_{\alpha}}+O\left(x _{\alpha}^{1/2}\right).
	\label{eq:AI_broken_density1_series}
\end{align}
As above, the exact calculation in Appendix~\ref{sec:Exact Expression for transition probability} gives the constant $\alpha$ of the first non-trivial term in Eq. (\ref{eq:AI_broken_density1_series}), where $\alpha \approx 0.12$.

The agreement of this expression with the numerical calculations is remarkable, as depicted in Fig. \ref{fig:Pb_6}. A comparison of Eq. (\ref{eq:Pb_near_ep}) to numerics for $\tau_{Q}<1.2$ and $\tau_{Q}<0.6$ shows good agreement. For larger $\tau_{Q}$, however, the agreement gradually decreases, which we attribute to the fact that start point $t_{i}=\tau_{Q} \sinh \alpha_{0}$ maybe beyond (less) the impulse regime $[-\hat{t},\hat{t}]$, so that the assumption that the initial evolution is an impulse or the end stage is adiabatic is violated. To avoid this, the start point $t_{i} \ll \hat{t}$ or $t_{f} \gg -\hat{t}$, i.e., the whole evolution within the two stages as described in Eq. ($\ref{Schematic_diagram_broken}$).

\section{summary}
\label{sec:summary}
We have shown that, based on the assumption of the Kibble-Zurek mechanism, the AI approximation can be generalized to and provides good quantitative predictions about the adiabatic dynamics of the non-Hermitian two-level Landau-Zener-like systems. The AI approximation is key to the KZ mechanism. It can be used to calculate excitations in the frozen region, where we use relative occupation instead of projections on the excited state in the Hermitian system. At first, during quenching, only in PT-symmetric systems, we find a KZ-like dependence of the topological defect density on the quench rate similar to the Hermitian Landau–Zener model. Then, when the entire evolution is a full non-Hermitian drive process, we find that the reciprocal of the absolute value of the complex energy spectrum can characterize the relaxation time of the system and then can distinguish between adiabatic and freezing regimes. The result is that the excitation of the KZ-like mechanism also predicts the density of defects in the freezing regimes of the PT-broken phase. 

We expect that the AI approximation can be generalized to arbitrary non-Hermitian two-level models, as well as many-body models such as the Ising model \cite{non_equilibrium4} and even incommensurate models such as the Aubry-André model \cite{non_equilibrium5,incommensurate2, AA1, AA2, AA3, AA4}. Finally, our results for slow quenching in the vicinity of EP can be directly verified experimentally: as in cold Fe$_{8}$ clusters \cite{LZ2, LZ_model2}, Mach-Zender interferometer \cite{KZ3}, and single-photon interferometry\cite{NH_EP5}.

\begin{acknowledgments}
We are grateful to Yiling Zhang, Xueping Ren, Yeming Meng, and Yue Hu for comments and valuable suggestions on the manuscript. This work is supported by NSFC Grant No. 11974053, 12174030.
\end{acknowledgments}

\appendix
\section{EXACT EXPRESSION FOR TRANSITION PROBABILITY}
\label{sec:Exact Expression for transition probability}
To test the predictions of the AI approximation for a PT-symmetric two-level system different from the classical Landau-Zener model and give the exact value of the constant $\alpha$, we consider in this section the dynamics problem induced by the Hamiltonian
\begin{equation}
	H(t)=\frac{1}{2}\left(\begin{array}{cc}
	(-1)^n \operatorname{sgn}(t)\left|\frac{t}{\tau_Q}\right|^{\frac{\eta}{1-\eta}} & 1 \\
	1-\delta & (-1)^{n+1} \operatorname{sgn}(t)\left|\frac{t}{\tau_Q}\right|^{\frac{\eta}{1-\eta}}
\end{array}\right),
\label{Hamiltonian_generalized}
\end{equation}
where $\eta \in(0,1)$ are constant parameters (when $\eta=1/2$, the system is simplified to the non-Hermitian Landau-Zener model (\ref{Hamiltonian_symmetry})). We want to give the lowest order exact expression of the exciting probability in a class of the non-Hermitian two-level systems described by the Hamiltonian (\ref{Hamiltonian_generalized}).

Firstly, when $n=0, \delta=2$ and assume that evolution occurs in the PT-symmetric regimes, e.g., in the interval $|t|>\tau_{Q}$ when $\eta=1/2$. Then express the time-depended wave function as
\begin{equation}
\begin{aligned}
	|\Psi(t)\rangle=& C_1(t) \exp \left(\frac{-i(1-\eta)|t|^{1 /(1-\eta)}}{2 \tau_Q^{\eta /(1-\eta)}}\right)|1\rangle \\
	&+C_2(t) \exp \left(\frac{i(1-\eta)|t|^{1 /(1-\eta)}}{2 \tau_Q^{\eta /(1-\eta)}}\right)|2\rangle,
\end{aligned}
\end{equation}
where the exponential is from $\mp i \int d t \operatorname{sgn}(t)\left|t / \tau_Q\right|^{\eta /(1-\eta)} / 2$. This wave function is reduced by the time-dependent Schrödinger equation into two first-order differential equations concerning C's
\begin{align}
	\dot{C}_1(t) &=\frac{1}{2i} \exp \left(\frac{i(1-\eta)|t|^{1 /(1-\eta)}}{\tau_Q^{\eta /(1-\eta)}}\right) C_2(t), \label{eq:dc1}
	\\
	\dot{C}_2(t) &=\frac{1}{2i} k \exp \left(\frac{-i(1-\eta)|t|^{1 /(1-\eta)}}{\tau_Q^{\eta /(1-\eta)}}\right) C_1(t). \label{eq:dc2}
\end{align}
(i) \textit{Time evolution from the ground state of $t_{i}\rightarrow -\infty$ to $t_{f} \rightarrow -t_{n}$}, here $t_{n}$ means evolution in the vicinity of EP: We want to integrate Eq. (\ref{eq:dc1}) from $-\infty$ to $-t_{n}$. The initial conditions are $C_{1}(-\infty)=1$, $C_{2}(-\infty)=0$. Simplification occurs when one assumes a quite fast transition, i.e., $\tau_{Q}\rightarrow 0$. Then obviously $C_{1}(t)=1+O(\tau_{Q}^{\beta})$ with some $\beta>0$. Putting such $C_{1}(t)$ into Eq. (\ref{eq:dc2}), we get
\begin{align}
C_2(-\gamma_{n}) &=\frac{1}{2 i} k \tau_Q^\eta \int_{-\infty}^{-\gamma_{n}} d \gamma \exp \left[-i(1-\eta)|\gamma|^{1 /(1-\eta)}\right] \notag\\
		    &+(\text{higher order terms in $\tau_{Q}$}),
\end{align}
where $-\gamma_{n}$ means the point near $-\gamma_{EP}$. $C_1(-\gamma_{n}) \approx 1$, which after some algebra results in 
\begin{align}
D_{2} &=\left|  \braket{2^{L}|\Psi(t_{f})}\right|^2=1-\left|  \braket{1^{L}|\Psi(t_{f})}\right|^{2}=\frac{1}{4}\left( \alpha\tau_{Q} \right)^{2\eta} \notag\\
	&+\left( \text{higher order terms in $\tau_{Q}$} \right)
   \label{eq:symmetry_density1},
\end{align}
where
\begin{small}
\begin{align}
\alpha&=\frac{\sqrt{11} \pi}{1728}  \left[72 \left(C\left(\frac{6}{5 \sqrt{\pi }}\right)-1\right) C\left(\frac{6}{5 \sqrt{\pi }}\right)\right]\notag\\
&-\frac{28}{1728}  \left[\left(S\left(\frac{6}{5 \sqrt{\pi }}\right)-1\right) S\left(\frac{6}{5 \sqrt{\pi }}\right)+11\right],
\end{align}
\end{small}
here $S$ and $C$ are Fresnel integrals, $C(z)=\int _0^z d t  \cos\left(\pi  t^2/2\right) $.

(ii) \textit{Time evolution from the ground state of $t_{i}\rightarrow t_{n}$ to $+ \infty$}: Since the initial function is $\frac{-i}{\sqrt{2}} \sqrt{\frac{\gamma_{0}}{\sqrt{\gamma_{0}^{2}-1}}+1} \ket{1^{R}}+ \frac{i}{\sqrt{2}} \sqrt{\frac{\gamma_{0}}{\sqrt{\gamma_{0}^{2}-1}}+1} \ket{2^{R}}$. For fast transition one has that $C_{1}(t)=\frac{-i}{\sqrt{2}} \sqrt{\frac{\gamma_{0}}{\sqrt{\gamma_{0}^{2}-1}}+1} + O(\tau_{Q}^{\beta})$, $C_{2}(t)=\frac{i}{\sqrt{2}} \sqrt{\frac{\gamma_{0}}{\sqrt{\gamma_{0}^{2}-1}}+1}+O(\tau_{Q}^{\delta})$, where $\beta, \delta>0$ are some constants. Integrating Eq. (\ref{eq:dc2}) from $\gamma_{0}$ to $\infty$, one gets
\begin{align}
	C_2(+\infty)&=\frac{i k \tau_Q^\eta }{2 \tanh{\frac{\pi}{4}}} \int_{\gamma_{0}}^{+\infty} d \gamma \exp \left[-i(1-\eta) \gamma^{1 /(1-\eta)}\right] \notag\\
	 		 &-\frac{i}{\sqrt{2}} \sqrt{\frac{\gamma_{0}}{\sqrt{\gamma_{0}^{2}-1}}+1}\notag\\
			 &+\left( \text{higher order terms in $\tau_{Q}$} \right),
\end{align}
which can be easily proved to lead to
\begin{align} 
	\mathcal{D}_{1}&=\left|  \braket{1^{L}|\Psi(t_{f})}\right|^2=1-\left|  \braket{2^{L}|\Psi(t_{f})}\right|^{2}=\frac{3}{10}-c_{1} (\alpha \tau_{Q})^{\eta} \notag\\
				&+\left( \text{higher order terms in $\tau_{Q}$} \right),
	\label{eq:symmetry_density3}
\end{align}
where $c_{1}=\sinh ^2\left(\frac{\pi }{4}\right) \text{sech}\left(\frac{\pi }{2}\right)$, and $\alpha=\frac{\pi \left(1+e^{\pi }\right)}{8 \left(e^{\pi /2}-1\right)^4} \left(1-2 S\left(\frac{\coth \left(\frac{\pi }{2}\right)}{\sqrt{\pi }}\right)\right)^2$.

(iii) \textit{Time evolution from $t_{i} \rightarrow -\infty$ to $t_{f} \rightarrow -\gamma_{EP}$ in the PT-broken regime:} To correctly calculate the density of topological defects, we need to introduce the relative population (\ref{eq:relative occupation}) instead of Eq. (\ref{eq:PT_density_1}). And we assume that the Hamiltonian is given by Eq. (\ref{Hamiltonian_generalized}) with $n=1/2, \delta=0$.

Because the parameter $n=1/2, \delta=0$ are different from the value of the PT-symmetric phase (\ref{Hamiltonian_symmetry}), the wave function is rewritten as
\begin{align}
	|\Psi(t)\rangle&= C_1(t) \exp \left(\frac{(1-\eta)|t|^{1 /(1-\eta)}}{2 \tau_Q^{\eta /(1-\eta)}}\right)|1\rangle \notag \\
	&+C_2(t) \exp \left(\frac{-(1-\eta)|t|^{1 /(1-\eta)}}{2 \tau_Q^{\eta /(1-\eta)}}\right)|2\rangle.
\end{align}
Bringing the above wave function into the Schrödinger equation, it can be reduced to
\begin{align}
	 \dot{C}_{1}(t)=&\frac{C_{2}(t)}{2i} \exp \left(\frac{-(1-\eta)|t|^{1 /(1-\eta)}}{\tau_Q^{\eta /(1-\eta)}}\right) \label{broken_dc1},\\
	 \dot{C}_{2}(t)=&\frac{C_{1}(t)}{2i} \exp \left(\frac{(1-\eta)|t|^{1 /(1-\eta)}}{\tau_Q^{\eta /(1-\eta)}}\right) \label{broken_dc2}.
\end{align}
Because the unnormalized initial wave function is $\left(i e^{\pi /2}, 1\right)$ we have $C_{1}(\gamma_{0})=i e^{\pi /2}$ and $C_{2}=1$. For fast transition, one has that $C_{1}(t)=i e^{\pi /2}+O(\tau_{Q}^{\beta})$, where $\beta>0$ is a constant. Integrating Eq. (\ref{broken_dc1}) and Eq. (\ref{broken_dc2}) from $-\infty$ to $-t_{n}$, one gets
\begin{align}
	 C_1(\gamma_{f})&=\frac{\tau_Q^\eta}{2 i} i e^{\pi /2} \int_{-\infty}^{-\gamma_{n}} d \gamma \exp \left[-(1-\eta) \gamma^{1 /(1-\eta)}\right] \label{broken_intc1} \notag\\
	 &+(\text{high order terms of}~\tau_{Q}),\\
	 C_2(\gamma_{f})&=\frac{\tau_Q^\eta}{2 i} \int_{-\infty}^{-\gamma_{n}} d \gamma \exp \left[(1-\eta) \gamma^{1 /(1-\eta)}\right]+1
	  \label{broken_intc2}\notag\\
	 &+ (\text{high order terms of}~\tau_{Q}).
\end{align}
which will easily lead to
\begin{align}
	\mathcal{D}_{r}&=\frac{\left|  \braket{1^{L}|\Psi(t_{f})}\right|^2}{\left|  \braket{2^{L}|\Psi(t_{f})}\right|^2 + \left|  \braket{1^{L}|\Psi(t_{f})}\right|^2} \notag\\
	&= \frac{1}{12} \left(6-\sqrt{11}\right)-\left( \alpha \tau_{Q}\right)^{\eta} \notag\\
	&+(\text{high order terms of}~\tau_{Q}),
	\label{eq:broken_density1}
\end{align}
where
\begin{small}
\begin{align}
	\alpha &\approx\frac{15 \pi}{7442} \left(\left(11+\sqrt{11}\right) \sqrt{6+\sqrt{11} }\sqrt{22 \left(47-12 \sqrt{11}\right) } \notag \right.\\
	&\left. \left(\text{erf}\left(\sqrt{\frac{3}{2}}\right)-\text{erf}\left(\frac{3 \sqrt{2}}{5}\right)\right)\left(\text{erfi}\left(\sqrt{\frac{3}{2}}\right)-\text{erfi}\left(\frac{3 \sqrt{2}}{5}\right)\right) \right),
	\label{eq:alpha_broken_3}
\end{align}
\end{small}
here	$\text{erf}, \text{erfi}$ are the Gauss error function $\operatorname{erf}(z)=\frac{2}{\sqrt{\pi}} \int_0^z e^{-t^2} d t$, $\operatorname{erfi}(z)=\frac{2}{\sqrt{\pi}} \int_0^z \mathrm{e}^{t^2} \mathrm{~d} t$ and the virtual error function respectively.

(iv) \textit{Time evolution from $t \rightarrow t_{n}$ to $t \rightarrow t_{f}$ in the PT-broken regime:} Because at $t\rightarrow \infty$, the positive energy eigenstate will dominate the occupation probability, and the exponential increase of the least dissipative state over time suppresses another eigenstate with negative imaginary energy, so it is required time truncation $t_{f}\approx \hat{t}$. Integration of Eq. (\ref{broken_dc1}) and Eq. (\ref{broken_dc2}) separately gives
\begin{align}
	 C_1(\gamma_{f})&=\frac{\tau_Q^\eta}{2 i} i e^{\pi /2} \int_{\gamma_{n}}^{\gamma_{f}} d \gamma \exp \left[-(1-\eta) \gamma^{1 /(1-\eta)}\right] +1
	 \label{broken_intc1} \notag\\
	 &+(\text{high order terms of}~\tau_{Q}),\\
	 C_2(\gamma_{f})&=\frac{\tau_Q^\eta}{2 i} \int_{\gamma_{n}}^{\gamma_{f}} d \gamma \exp \left[(1-\eta) \gamma^{1 /(1-\eta)}\right]+i e^{\pi /2} \label{broken_intc2}\notag \\ 
	 &+(\text{high order terms of}~\tau_{Q}),
\end{align}
Then the relative occupation leads to the following prediction:
\begin{align}
	\mathcal{D}_{r}&=\frac{\left|  \braket{2^{L}|\Psi(t_{f})}\right|^2}{\left|  \braket{2^{L}|\Psi(t_{f})}\right|^2 + \left|  \braket{1^{L}|\Psi(t_{f})}\right|^2} \notag\\
	&= \frac{1}{12} \left(6-\sqrt{11}\right)+\left( \alpha \tau_{Q}\right)^{\eta} \notag\\
	&+(\text{high order terms of}~\tau_{Q}),
	\label{eq:broken_density3}
\end{align}
with $\alpha$ is same as Eq. (\ref{eq:alpha_broken_3}).

\section{EXACT SOLUTIONS OF THE NON-HERMITIAN LANDAU-ZENER SYSTEM}
\label{appendixa}
In this part, we want to present the exact solution to the dynamics of the non-Hermitian two-level model from the Landau-Zener theory. The model that we consider is the same as the Eq. (\ref{Hamiltonian_symmetry}), described by the ordinary differential equations for probability amplitudes $c_{1}$ and $c_{2}$,
\begin{align}
i \frac{d}{d t} c_{1}(t) &=\frac{1}{2} \gamma(t) c_{1}(t)+\frac{1}{2} \nu c_{2}(t), \notag\\
i \frac{d}{d t} c_{2}(t) &=\frac{1}{2} \nu(1-\delta) c_{1}(t)+-\frac{1}{2} \gamma(t) c_{2}(t).
\label{differential equations}
\end{align}
We can get the second-order equation of $c_{2}(t)$ by decoupling the above equation by differentiating it again
\begin{align}
	c_{2}^{''}(t)+\frac{1}{4}\left(k+\frac{t^{2}}{\tau_{Q}^{2}}-\frac{2 \dot{1}}{\tau_{Q}}\right) c_{2}(t)=0.
	\label{differential equation of C1}
\end{align}
Then a solution satisfying the initial condition $\left|c_{1}(t \rightarrow-\infty)\right|^{2}=0, \quad\left|c_{2}(t \rightarrow-\infty)\right|^{2}=1$ is given by 
\begin{equation}
\text{c}_{2}(t)= a D_{-\frac{1}{4} i k \text{$\tau_{Q}$}-1}\left(\frac{ e^{i \pi/4} t}{\sqrt{\text{$\tau_{Q}$}}}\right)+b D_{-\frac{1}{4} i k \text{$\tau_{Q}$}-1}\left(-\frac{ e^{i \pi/4} t}{\sqrt{\text{$\tau_{Q}$}}}\right),
\end{equation}
the general solution can be expressed in terms of independent Weber function $D_{\nu}(z)$, where the $\nu=\frac{1}{4} i k \tau_{Q}$, $z=-\frac{ e^{i \pi/4} t}{\sqrt{\tau_{Q}}}$. Then, the time-dependent wave function can be obtained from the Schr\"{o}dinger equation
\begin{equation}
\begin{aligned}
\begin{aligned}
|\Psi(t)\rangle&= \frac{2 i}{1-\delta}\left[\partial_{t}-\frac{i t}{2 \tau_{Q}}\right]\left[a D_{-\nu-1}(i z)+b D_{-\nu-1}(-i z)\right]|1\rangle \\
&+\left[a D_{-\nu-1}(i z)+b D_{-\nu-1}(-i z)\right]|2\rangle.
\end{aligned}
\label{time wave}
\end{aligned}
\end{equation}
The constants $a$ and $b$ are to be found from the initial values $c_{1}(t_{i})$ and $c_{2}(t_{i})$
\begin{equation}
\begin{aligned}
&\begin{aligned}
a=& \frac{\Gamma(1-\nu)}{\sqrt{2 \pi}}\left[D_{\nu-1}\left(-z_i\right) c_1\left(t_i\right)+2\sqrt{\tau_{Q}} e^{i \pi / 4} D_\nu\left(-z_i\right) c_2\left(t_i\right)\right],
\end{aligned}\\
&b=\frac{\Gamma(1-v)}{\sqrt{2 \pi}}\left[D_{v-1}\left(z_i\right) c_1\left(t_i\right)-2\sqrt{\tau_{Q}} e^{i \pi / 4} D_v\left(z_i\right) c_2\left(t_i\right)\right] \text {. }
\end{aligned}
\label{eq:symmetry_a,b}
\end{equation}

(i) \textit{Exact solution of non-Hermitian LZ problem when evolution starts at $t_{i}\rightarrow -\infty$}, i.e., $\ket{\Psi(-\infty)}\sim \ket{1^{R}}$. The Hamiltonian is given by Eq. (\ref{Hamiltonian_symmetry}), together with the $c_{2}(t)$ and initial condition 
yields $a=0, b=e^{ (1-\delta) \pi \tau_{Q}/16}$. Substituting them into Eq. (\ref{time wave}) results in
\begin{align}
|\Psi(t)\rangle&=\frac{2 e^{i 3 \pi/ 4}e^{\frac{(1-\delta) \pi \tau_{Q}}{16}} D_{-\frac{1}{4} i k \tau \mathrm{q}}\left(-\frac{ e^{i \pi/4} t}{\sqrt{\tau q}}\right)}{k \sqrt{\tau \mathrm{q}}}|1\rangle  \notag\\
&+e^{\frac{(1-\delta) \pi \tau_{Q}}{16}} D_{-\frac{1}{4} i k \tau_{Q}-1}\left(-\frac{ e^{i \pi/4} t}{\sqrt{\tau_{Q}}}\right)|2\rangle.
\end{align}
And the density defect can be calculated by the $\mathcal{D}_{\uparrow}$ one obtains the excitation probability of the system at $t_{f}$ in the form of Eq. (\ref{eq:symmetry_density1}).

(ii) \textit{Exact solution of LZ problem when evolution starts from a ground state at PT-broken regime}: from $t_{i}\rightarrow t_{n}$ to $t_{f} \rightarrow \infty$. Combining the initial condition $c_{1}(t_{i})= \frac{-i}{\sqrt{2}} \sqrt{\frac{\gamma_{0}}{\sqrt{\gamma_{0}^{2}-1}}+1}, c_{2}(t_{i})=\frac{i}{\sqrt{2}} \sqrt{\frac{\gamma_{0}}{\sqrt{\gamma_{0}^{2}-1}}+1}$ and Eq. (\ref{eq:symmetry_a,b}), the constants $a$ and $b$ can be obtained
\begin{align}
a=& \frac{1}{\sqrt{2 \pi }}+\frac{\left( 6+\sqrt{11}\right)e^{\frac{i \pi }{4}}}{20} \sqrt{\text{$\tau$}_{Q}}, \notag\\
b=&\frac{1}{\sqrt{2 \pi }}-\frac{ \left( 6+\sqrt{11}\right)e^{\frac{i \pi }{4}}}{20} \sqrt{\text{$\tau$}_{Q}} \text {. }
\end{align}
From this, it is easy to prove that when $t_{f} \rightarrow \hat{t}$, the modulus of  the projection of the system on state $\ket{2^{L}}$ is equal to Eq. (\ref{eq:symmetry_density3}).

However, the PT-broken regime is too small to see the adiabatic-impulse transition. The Hamiltonian is replaced by the case of $n=1/2, \eta=1/2$ in the Eq. (\ref{Hamiltonian_symmetry}), and the ordinary differential equations are
\begin{align}
i \frac{d}{d t} d_{1}(t) &=\frac{1}{2} i\gamma(t) d_{1}(t)+\frac{1}{2} \nu d_{2}(t), \notag\\
i \frac{d}{d t} d_{2}(t) &=\frac{1}{2} \nu d_{1}(t)-\frac{1}{2} i\gamma(t) d_{2}(t),
\label{differential equations of another}
\end{align}
for easily, the $\delta=0$. And the process to get the $d_{1}$ and $d_{2}$ is the same as the Eq. (\ref{differential equation of C1}), but with $\nu=\frac{\tau_{Q}}{4}, z=\frac{t}{\sqrt{\tau_{Q}}}$. By combining this observation with $\eta=1/2$ version of Eq. (\ref{broken_intc2}) one gets
\begin{align}
	|\Psi(t)\rangle&= i \left[2 \partial_{t}+\frac{ t }{ \tau_{Q}}\right]\left[a D_{-\nu-1}(i z)+b D_{-\nu-1}(-i z)\right]|1\rangle \notag\\
&+\left[a D_{-\nu-1}(i z)+b D_{-\nu-1}(-i z)\right]|2\rangle.
\label{broken_psit}
\end{align}
Then the constants $a$ and $b$ from Eq. (\ref{broken_psit}) turn out to be equal to
\begin{align}
  a=&\frac{1}{c} \left( \sqrt{\tau_{Q}} D_{-\nu-1}\left(i z_{i}\right) c_{1}(t_{i})-2 D_{-\nu}\left( i z_{i}\right) c_{2}(t_{i}) \right), \notag\\
  b=&\frac{1}{c} \left(\sqrt{\tau_{Q}} D_{-\nu-1}\left(-z_{i}\right) c_{1}(t_{i})+2D_{-\nu}\left(-i z_{i}\right) c_{2}(t_{i}) \right) \text {. }
\label{eq:broken_a,b}
\end{align}
where $c=2 D_{-\nu -1}\left(i z_i\right) D_{-\nu }\left(-i z_i\right)+2 D_{-\nu -1}\left(-i z_i\right) D_{-\nu }\left(i z_i\right).$ 

(iii) \textit{Exact solution of LZ problem when evolution starts from a ground state at PT-broken regime}: from $t_{i}\rightarrow -\infty$ to $t_{f} \rightarrow -\hat{t}$. The initial state is $\ket{\Psi(0)}=\ket{1^{R}}$. Combining the initial condition $c_{1}(t_{i}),c_{2}(t_{i})$ and Eq. (\ref{broken_psit}), the constants $a$ and $b$ can be obtained
\begin{align}
a&= \frac{1}{8 \sqrt{1+e^{\frac{5 \pi }{2}}} \pi} \left(2 \sqrt{2 \pi}\left(2 e^{5 \pi / 4}+\operatorname{csch}\left(\frac{5 \pi}{4}\right)\right) \right.\notag\\
&\left.+2 i \pi \sqrt{\tau_Q}+\left(-3+e^{5 \pi / 2}\right) \sqrt{\frac{\pi}{2}} \operatorname{cosh}\left(\frac{5 \pi}{4}\right) \tau_Q \right)\notag\\
b&= \frac{1}{8 \sqrt{1+e^{\frac{5 \pi }{2}}} \pi} \left(2 \sqrt{2 \pi}\left(2 e^{5 \pi / 4}+\operatorname{csch}\left(\frac{5 \pi}{4}\right)\right) \right.\notag\\
&\left.-2 i \pi \sqrt{\tau_Q}+\left(-3+e^{5 \pi / 2}\right) \sqrt{\frac{\pi}{2}} \operatorname{cosh}\left(\frac{5 \pi}{4}\right) \tau_Q \right). \notag\\
\end{align}
Note that due to the non-Hermitian non-unitary evolution, the time needs to be truncated, and it can be proved that when $t_{i} \rightarrow -\cosh{\frac{5\pi}{4}} \tau_{Q}$, is shown in Fig. (\ref{fig:Pb_6}), the relative occupation is equal to (\ref{eq:broken_density1}).

(iv) \textit{Exact solution of LZ problem when evolution starts from a ground state at $t_{i} \rightarrow \text{arccosh} (\theta_{0}) \tau_{Q}$}: When $\theta_{0}=0.2\pi$, the initial state is $\ket{\Psi(0)}=\left(i/\sqrt{1+e^{\frac{2 \pi }{5}}},1/\sqrt{1+e^{-\frac{2 \pi }{5}}}\right)$. The constants $a$ and $b$ from Eq. (\ref{eq:broken_a,b}) are proved to be equal to
\begin{equation}
\begin{aligned}
&\begin{aligned}
a=&\sqrt{\frac{\pi }{2}} \left(-\frac{i \left(\sqrt{11}+6\right)}{20} \sqrt{\tau_{Q}}+\frac{25-3 \left(\sqrt{11}+3\right)}{25 \sqrt{2 \pi }}\tau_{Q}\right),
\end{aligned}\\
&b=\sqrt{\frac{\pi }{2}} \left(-\frac{i \left(\sqrt{11}+6\right)}{20}  \sqrt{\tau_{Q}}-\frac{25-3 \left(\sqrt{11}+3\right)}{25 \sqrt{2 \pi }}\tau_{Q}\right) \text {. }
\end{aligned}
\end{equation}
Then, simple calculation directly leads to expression Eq. (\ref{eq:broken_density3}).
\newpage
\nocite{*}
\bibliography{NH_LZ}
\end{document}